\documentclass[graybox]{svmult}

\usepackage{mathptmx}       
\usepackage{helvet}         
\usepackage{courier}        
\usepackage{type1cm}        
\usepackage{makeidx}         
\usepackage{graphicx}        
\usepackage{multicol}        
\usepackage[bottom]{footmisc}
\usepackage{hyperref}		%

\usepackage{color}
\definecolor{keywordcolor}{rgb}{0,0,0} 
\providecommand{\keyword}[1]{\index{#1}{\color{keywordcolor}#1}}

\makeindex             

\begin{document}

\title*{Resonant excitation and photon entanglement from semiconductor quantum dots}
\author{Ana Predojevi\'{c}}
\institute{Ana Predojevi\'{c} \at Institut f\"ur Experimentalphysik, Universit\"at Innsbruck, Technikerstr. 25, 6020 Innsbruck, Austria \email{ana.predojevic@uibk.ac.at}}
%
%
\maketitle


\abstract{In this chapter we review the use of semiconductor quantum dots as sources of quantum light. Principally, we focus on resonant two-photon excitation, which is a method that allows for on-demand generation of photon pairs. We explore the advantages of resonant excitation and present a number of measurements  that were made in this excitation regime. In particular, we cover the following topics: photon statistics, coherent manipulation of the ground-excited state superposition, and generation of time-bin entangled photon pairs. 
\newline\indent
}

\section{Introduction}
\label{sec:introduction}
The field of photonic quantum information needs novel, highly efficient, and \keyword{deterministic sources of single photons} and \keyword{entangled photon pairs}.  The principal applications of these sources include quantum networks \cite{Briegel1998,Duan2001, Kimble2008, Simon2003} and linear optical quantum computing \cite{Knill2001}. 
In particular, quantum light is needed to transfer information in procedures like teleportation \cite{Bouwmeester1997} and entanglement swapping \cite{Zukowski1993}; the photons are employed as flying qubits that interconnect the nodes of a quantum network, or to run a quantum processor using the methods of linear optical quantum computing.

The vast majority of today's quantum information experiments use single photons and entangled photon pairs that are generated in a process of parametric down-conversion. While this method still stands as the most versatile and successful, semiconductor \keyword{quantum dot} devices are developed because of their potential to deliver a source that is brighter and more reliable but also can be easily integrated within a semiconductor optical circuit.  While the initial interest in quantum dots in general was more oriented towards semiconductor and material physics and even chemistry, today's semiconductor quantum dot devices are also very often designed for the purposes of quantum information processing. Namely, similar to atoms quantum dots possess discrete energy structure and therefore a valuable asset of an intrinsic \keyword{sub-Poissonian distribution} of the emitted photons. Due to their atom-like energy structure quantum dots can emit single photons \cite{Michler2000} but their range of application does not end there. They can also deliver pairs of photons, emitted in a temporary ordered cascade. 
In addition to their potential to be used as sources of photons, quantum dots can also take the role of a quantum memory. In particular, the quantum dot potential can also trap single carriers (electrons and holes) and the spin on such a carrier can encode a quantum bit (see the chapter by McMahon \& De Greve). 


In Section \ref{sec:on-demand} of this chapter we will give the basic specifications that a photon source should fulfil in order to be used in a specific application. Also we will introduce an excitation method that can resonantly create pairs of photons from a quantum dot. We will address this problem from both an experimental and a theoretical point of view. In addition, we will review the use of quantum dots to generate polarization entangled photon pairs. In Section \ref{sec:coherent and time} we will present measurements that exploit the use of resonant excitation. In particular, we will address the coherent control, the effects resonant excitation has on the photon statistics of the emitted light, and finally the generation of time-bin entangled photon pairs emitted by a single semiconductor quantum dot.

\section{On-demand generation of photon pairs using single semiconductor quantum dots}
\label{sec:on-demand}

There are a number of applications that need or benefit from single photons and entangled photon pairs. These include linear optical quantum computing, long distance quantum communication, and up to some level quantum cryptography. 
Though certain tasks can be performed in a probabilistic manner or even override the use of single photons, the optimum performance and minimal overhead are very often achieved using a deterministic photon source.

Probably the most straightforward example is linear optical quantum computing \cite{Kok2007}. Photonic quantum computing using linear elements, as proposed in the seminal paper by Knill, Laflamme, and Milburn \cite{Knill2001}, is a method to realise a quantum processor. The proposal in its original form assumes an ideal single photon source. Posteriorly, a scheme was shown \cite{Varnava2008} that allows linear optical quantum computation if the overall efficiency (source $\times$ detector) is higher then 2/3. Nevertheless, to achieve gates outside the post-selection basis\footnote{The inefficiency of sources and detectors are commonly bridged using post-selection.} one needs much higher photon generation probability combined with a very low probability for emission of more than one photon \cite{Jennewein2011}. Concerning the use of photon sources within quantum networks it is harder to define an efficiency threshold because it would depend on the specific application\footnote{For example for a a complex task like distributed quantum computing the threshold will be different than for the simplest form of communication between two network nodes.}. Furthermore, quantum networks depends on many additional parameters like the efficiencies of state mapping or generation of atom-photon entanglement. Regardless the specific application, a deterministic photon source would surely increase the information transfer rate. On the other hand, multi-photon contribution, which we will show in continuation is greatly reduced in quantum dots under resonant excitation, has a negative effect on long-distance entanglement distribution \cite{Sangouard2007} as well as quantum key distribution \cite{Lutkenhaus2000}. 


%


\subsection{Quantum dots and polarization entanglement}

\begin{figure}[b]
\centering
\includegraphics[scale=0.75]{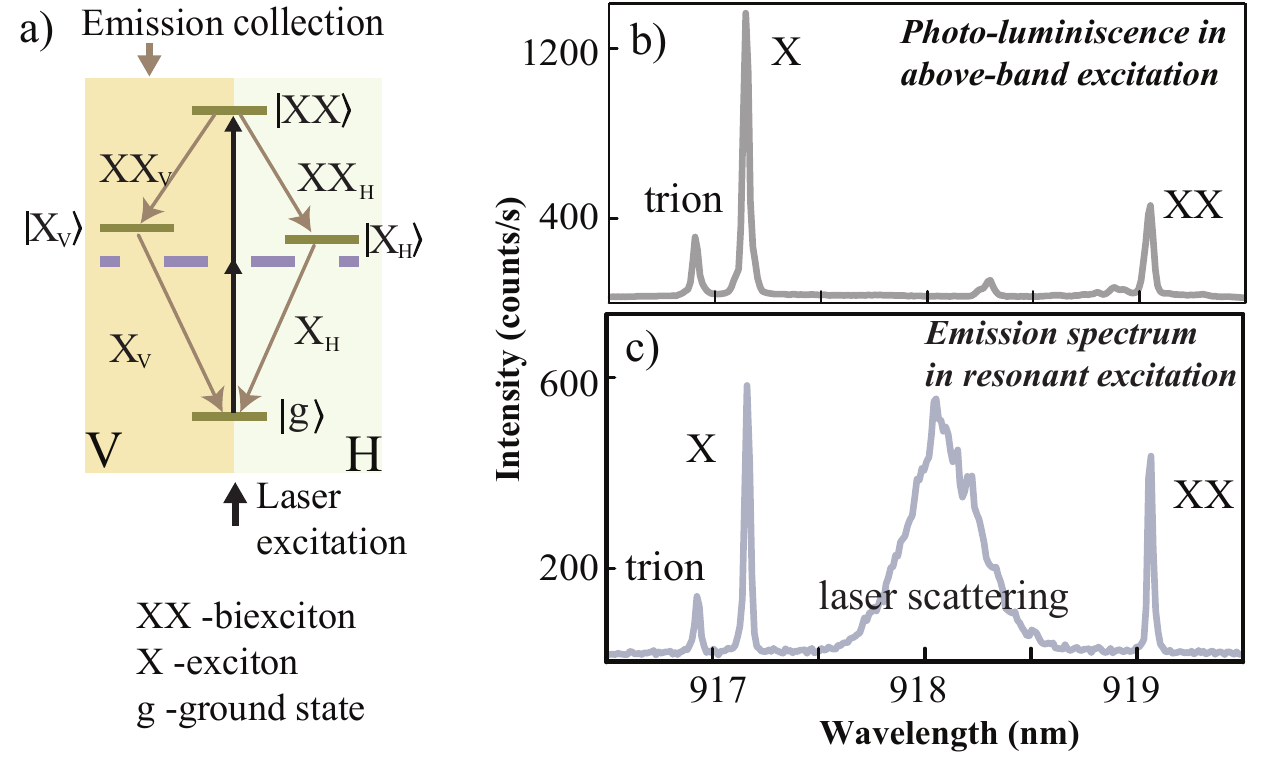}
\caption{(a) Quantum dot energy scheme. Fine structure splitting is the energy difference between two exciton levels $|X_{H}-X_{V}|$. In the process of two-photon resonant excitation a pulsed laser with the the energy $E_{laser} = (| XX\rangle-|g\rangle)/2$ (shown as black arrow pointing upwards) coherently couples the ground ($ |g\rangle$) and the biexciton ($|XX\rangle$) states through a virtual level (dashed line). Biexciton recombination takes place through the intermediate exciton states ($|X_{H,V} \rangle$) emitting biexciton ($XX_{H,V}$) and exciton ($X_{H,V}$) photons. (b) Photo-luminescence emitted by a single quantum dot. The excitation laser of wavelength close to 630~nm generates a reservoir of carriers that via multiple phonon scattering occupy the quantum dot levels. Here, we observe three emission lines corresponding to exciton (X), biexciton (XX), and trion.  (c) Emission spectrum obtained in two-photon resonant excitation. The laser energy is half way between exciton and biexciton and not resonant to any of these transitions.}
\label{fig:2p}      
\end{figure}

Before entering the topic of resonant excitation, we will briefly review the use of quantum dots to generate \keyword{polarization entangled photon pairs}. This system has been proposed to be capable of delivering entangled photon pairs \cite{Benson2000} through the use of a biexciton-exciton (XX-X) photon cascade. In particular, once the quantum dot potential has trapped two electron-hole pairs (biexciton) the system decays to the ground state via an emission of the temporally ordered \keyword{photon cascade}. This decay can happen via two different paths that give photons with orthogonal polarizations. If these decay paths are indistinguishable the emitted pair of photons are entangled in polarization. Unfortunately this scheme is not straightforward to accomplish due to a geometrical anisotropy of quantum dots that is growth typical and almost unavoidable and that makes the intermediate exciton state split. This splitting (also known as the fine structure splitting - FSS, Fig.\ref{fig:2p}a) causes the two decay paths to be distinguishable. In other words, once the first photon of the cascade is emitted the system is projected in a superposition of the two 
exciton levels. This superposition evolves in time and therefore averages the phase of the emitted state \cite{Stevenson2008}, which causes the measured level of entanglement to be reduced with increasing exciton splitting.

There have been many attempts to overcome the problem of the fine structure splitting. The initial results were focused on picking the dots with the lowest splitting \cite{Young2006} or employing optical cavities to filter out a narrow indistinguishable spectral region \cite{Akopian2006}. In \cite{Muller2009} it was demonstrated that the optical Stark effect can be used to generate energy degenerate photons. 
Here, the Stark shift was used to tune the energy of the horizontally polarized exciton and make it degenerate with the vertically polarized one. Another specific approach was shown in \cite{Dousse2010} where a quantum dot was placed in a system of two strongly coupled micro-pillar cavities. Though the quantum dot itself was weakly coupled to the individual micro-pillars the overall effect was an enhanced emission of both exciton and biexiton photon. This resulted in the immediate emission of the exciton photon after the emission of the biexciton one. Under such experimental conditions the cascade pair is emitted faster than the phase of the generated state, due to the fine structure splitting, could have evolved.
 
Both fine structure splitting as well as the energies of the exciton and the biexciton can be tuned using an electric field \cite{Ghali2012}, a magnetic field \cite{Stevenson2006a}, or the material strain. In that respect one should point out the method shown in \cite{Trotta2014} where both electric field and the material strain were employed simultaneously. This approach allows for modification and removal of the fine structure for any randomly chosen quantum dot.
 
The origin of the exciton level splitting is growth induced, therefore, a number of experiments were demonstrated where the splitting was reduced via a modification of the growth method. In \cite{Kuroda2013} was explored  an alternative method of dot self assembly, namely, the droplet epitaxy on (111)A substrates. Another approach is the growth of so-called pyramidal quantum dots \cite{Juska2013} that apart from high geometrical symmetry can provide control of the position where the dot is grown. The use of quantum dots embedded in nanowires \cite{wires} is another growth related method that shares some similarities with the pyramidal quantum dots, in particular, intrinsic symmetry and the control of the position of the emitter. Here, it was proposed \cite{Singh2009} that the geometrical symmetry of the nanowires will condition the symmetry of the quantum dots embedded within. 

At this point it is also important to mention that all above-given results were achieved in above-band excitation.  The use of two-photon resonant excitation, in detail explained in continuation, has been shown to improve the degree of entanglement \cite{Muller2014}. 

There are different parameters that characterize two-qubit entanglement \cite{Guehne2004}; some rather being indicators (like fidelity $> 0.5$ to the maximally entangled state) and others being measures of entanglement (concurrence $> 0$ \cite{James2001}). While there are alternative methods to estimate the fidelity, to obtain a value for concurrence one needs to perform state tomography \cite{James2001}.  The results of the experiments that performed state tomography are summarized in Table \ref{tab:ent}.


%

\begin{table}
\caption{Characterization of the achieved polarization entanglement given for the experiments where both concurrence and fidelity were reported. The indicator of the nonlocality of the entanglement measurement - the violation of the Bell inequality - were reported in \cite{Kuroda2013} and \cite{Trotta2014}}
\label{tab:ent}       

\begin{tabular}{p{1.8cm}p{2.3cm}p{2.3cm}p{2.3cm}p{2.2cm}}
\hline\noalign{\smallskip}
 & \scriptsize{Young  (2006) } \cite{Young2006} &\scriptsize{Juska (2013)} \cite{Juska2013} &\scriptsize{Trotta  (2014)} \cite{Trotta2014} &\scriptsize{Huber (2014)} \cite{wires} \\
\noalign{\smallskip}\svhline\noalign{\smallskip}
\scriptsize{concurrence} & 0.44(3)  & 0.16(2) & 0.75(2)  & 0.57(6) \\
\scriptsize{fidelity} & 0.70(2)  & 0.58(3) & 0.82(4)  & 0.76(2) \\
\noalign{\smallskip}\hline\noalign{\smallskip}
\end{tabular}
\end{table}

\subsection{Resonant excitation}

A photon generation device employed in quantum information processing tasks must achieve a high success probability to produce a single photon. In atom-like systems such a behaviour is achievable by means of coherent population inversion. Likewise, the discrete energy structure of quantum dots makes this system suitable for driving such a process. 
  
On the other hand, despite the favourable energetic structure it is hard to achieve resonant excitation in semiconductor embedded quantum dots. The first, and most important reason is the excess \keyword{laser scattering} that is hard to distinguish from the single photon signal emitted by the quantum dot. Therefore, the traditional way to excite quantum dots is \keyword{above-band excitation}. Here, one uses a laser with an energy higher than any transition in the quantum dot. This laser creates a multitude of carriers in the vicinity of the quantum dot that can be probabilistically trapped in the quantum dot potential. This process is very nicely illustrated in the Fig.4 of the chapter by Schneider, Gold, Lu, H\"ofling, Pan \& Kamp. While it is possible to both saturate the quantum dot transitions and to achieve very high single photon count rates, the probabilistic nature of this process reduces the suitability of such a source for quantum information protocols. Another negative feature of the above-band excitation is related to how exactly the quantum dot levels are populated. Namely, biexciton photons will be created once the exciton level has been saturated and, therefore, the saturation of the biexciton level itself demands a very large number of carriers in the quantum dot vicinity. Such an experimental configuration is very unfavourable because it promotes the dephasing of the quantum dot levels due to the electric field fluctuations and causes poor photon statistics properties due to processes like carrier re-capture \cite{Peter2007}.

\keyword{Two-photon resonant excitation} of the biexciton \cite{Jayakumar2013} is an experimental implementation that simultaneously solves both problems: laser scattering and probabilistic generation of photon pairs. Here, one exploits the biexciton binding energy in order to drive the quantum dot system using a virtual resonance that is placed halfway in energy between the exciton and biexciton (see Fig.\ref{fig:2p}a) and therefore is not resonant to any of them. The photo-luminescence obtained in above-band excitation of the quantum dot is shown in Fig. \ref{fig:2p}b. There we can observe the lines of the quantum dot emission. For comparison, the emission spectrum under resonant excitation is shown in Fig. \ref{fig:2p}c. This spectrum shows an additional line coming from the scattered excitation laser light. The physical basis of the phenomenon we exploit here, the \keyword{biexciton binding energy}, is the Coulomb interaction present when two electron-hole pairs are trapped inside the quantum dot potential. As the first pair of carriers recombine and the biexciton photon is emitted the energy levels in the quantum dots will change and the second photon to be emitted (exciton photon) will not have the same energy as the biexciton photon. Therefore, we always observe the exciton and biexciton emission as two energetically well separated lines.

It is important to say that the two-photon approach to excite quantum dots is not new, nevertheless, it is quite challenging to apply this method on III-V quantum dots. The previous works \cite{Flissikowski2004} addressed  II-VI quantum dots that have much larger biexciton binding energy (the difference between the exciton and the biexciton line can be of even more than 10~nm) but have very unfavourable optical properties; they emit photons in the blue and green spectral range that are, due to losses in the optical fibres, not very suitable for quantum communication.  The values for the energy difference between biexciton and exciton lines in III-V quantum dots are in the region of 1-2~nm. Therefore, these systems demand a more thoughtful approach to reduce the laser scattering. The early works on III-V quantum dots \cite{Stufler2006} showed the signatures of resonant excitation, like for example \keyword{Rabi oscillations}, but only in photo-current measurements and not in the optical signal. The first optical measurements \cite{Jayakumar2013} showed Rabi oscillations as well as \keyword{Ramsey interference} measurements, while in \cite{Muller2014} it was also shown that resonant excitation can improve the degree of photon entanglement. 

Depending on the sample structure and the amount of power needed to excite the quantum dot, the resonant excitation of the biexciton might not be sufficient to fully suppress the laser scattering. Here, we will name two methods to additionally reduce the amount of laser scattering\footnote{In addition, one of the simplest approaches to minimize the laser scattering is the method of crossed polarisers \cite{Vamivakas2009}. Here, the excitation laser is horizontally polarized (Fig. \ref{fig:2p}a), while the emission is collected only from the vertically polarized ($XX_{V}-X_{V}$) cascade. 
}: sample/excitation geometry and design of pulse-bandwidth. The choice of the sample structure and the corresponding geometry of excitation can greatly reduce excess laser scattering. The method of orthogonal propagation paths was first shown in resonant excitation of a single exciton \cite{Muller2007}. The schematic of the excitation used in \cite{Muller2007} is shown in chapter: Schneider, Fig.~4d. Here, the excitation laser is directed onto the cleaved edge of the sample via an optical fibre that is brought to a distance of a few microns from the sample. This method was also used in  \cite{Jayakumar2013, Muller2014} with a difference that the laser light was focused onto the cleaved edge of the sample using an objective. 
 
\keyword{Micro-cavity quantum dot} samples, where the cavity extends all the way to the edges of the sample, are highly suitable structures for the implementation of this type of excitation geometry. Namely, the excitation laser is here focused onto the sample from the side, see Fig.\ref{fig:levels}a, so that the sample distributed Bragg reflector (DBR) structure acts as a waveguide for the laser light. The quantum dot emission is collected from the top using a high numerical aperture objective. 
For example, the specific sample used in \cite{Jayakumar2013} contained self-assembled InAs quantum dots of low density (approximately 10 per $\mu m^2$) 
that were embedded in a 4$\lambda$ thick, distributed Bragg reflector microcavity consisting of 15.5 lower and 10 upper $\lambda$/4 thick DBR layer pairs of AlAs and GaAs. The cavity mode was resonant at $\lambda$ = 920 nm. 
The results presented in \cite{Muller2014} were obtained using a sample with a $\lambda$ thick cavity that had far fewer upper-reflector DBR pairs.
.

As mentioned above, the spacing between the exciton and the biexciton line in III-V quantum dots is about 1-2~nm. This value can vary significantly even within the same quantum dot sample. Therefore the flexibility in choice of the excitation laser bandwidth is crucial for this application. There exist quite costly solutions for this problem, like for example lasers with variable pulse length. Nevertheless, in both \cite{Jayakumar2013} and \cite{Muller2014} it was shown that a combination of a short pulse laser (around 2~ps) and a pulse stretcher can fulfil both the variable bandwidth requirement as well as the need to fine tune the wavelength of the pulses. With respect to the design of the pulse stretcher special care should be given to the pulse chirp \cite{Muller2014}.

\subsection{Theoretical description of the two-photon excitation process}

In order to gain better understanding of the problem we introduce here a theoretical model of a three-level system subjected to the resonant two-photon excitation. The approach we present is very well known from atomic physics and describes resonant two-photon driving of a discrete-energy system in the presence of \keyword{level dephasing}. 

\begin{figure}[p]
\centering
\includegraphics[scale=1.0]{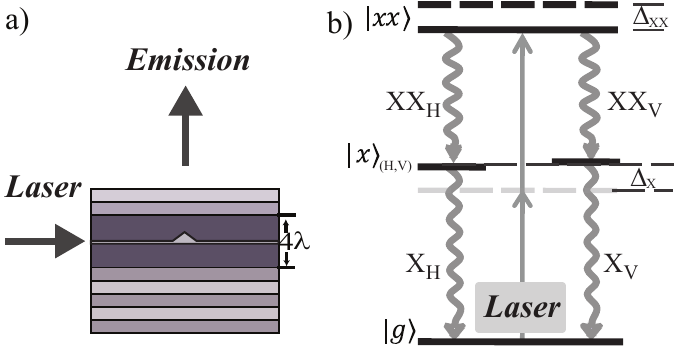}
\caption{(a) Schematics of the excitation geometry. The laser light is directed from the side onto the cleaved edge of the sample. Here, the 4$\lambda$ thick structure of the planar micro-cavity acts as a wave-guide for the excitation light. The quantum dot emission is collected from the top. (b) Energy level scheme for two-photon excitation of a biexciton. The excitation laser light coherently couples the ground ($ |g\rangle$) and the biexciton ($|xx\rangle$) states via the virtual level in two-photon resonance (dashed gray line). Biexciton recombination takes place through the intermediate exciton states ($|x_{H,V} \rangle$) emitting biexciton ($XX_{H,V}$) and exciton ($X_{H,V}$) photons. The energy difference between the exciton level and the two-photon virtual resonance is denoted as $\Delta_{x}$. For driving the two-photon transition off-resonantly we define the detuning $\Delta_{xx}$.}
\label{fig:levels}       
\end{figure}

The levels involved are the ground ($├ |g\rangle$), exciton state ($├ |x\rangle$), and biexciton state ($├ |xx\rangle$). The level scheme is shown in Fig.\ref{fig:levels}. The energy differences between ground state and exciton state, and between exciton state and biexciton state are not equal due to the  biexciton binding energy. 
 This electronic configuration allows for a two-photon excitation process where the pump laser is not resonant to any of the single photon transitions, while the two-photon process is resonant. To describe this system we can use the Hamiltonian of the following form:
\begin{equation}
H=\frac{\hbar\Omega_{1}\left(t\right)}{2}(\sigma_{g,x}+\sigma_{g,x}^{\dagger})+\frac{\hbar\Omega_{2}\left(t\right)}{2}(\sigma_{x,xx}+\sigma_{x,xx}^{\dagger})+\hbar\sigma_{x,x}(\Delta_{x}-\Delta_{xx})-2\hbar\sigma_{xx,xx}\Delta_{xx}
\end{equation}


Here, $\Omega_{l}(t)$, $\textit{l}$=1,2 is the Rabi frequency of the pump laser driving both single photon transitions. The transition operators and projectors are given as $\sigma_{i,j}=|i\rangle\langle j┤| $.  The energy difference between the virtual level of the two-photon transition and the exciton energy is $ \Delta_{x}$. This energy difference can also be seen as the laser detuning in a process of a single photon resonant excitation that drives the exciton state.  
To drive the two-photon transition off-resonantly we define the detuning $\Delta_{xx}$, the difference between the two-photon virtual resonance and the energy of the laser driving the system.

\begin{figure}[p]
\centering
\includegraphics[scale=.58]{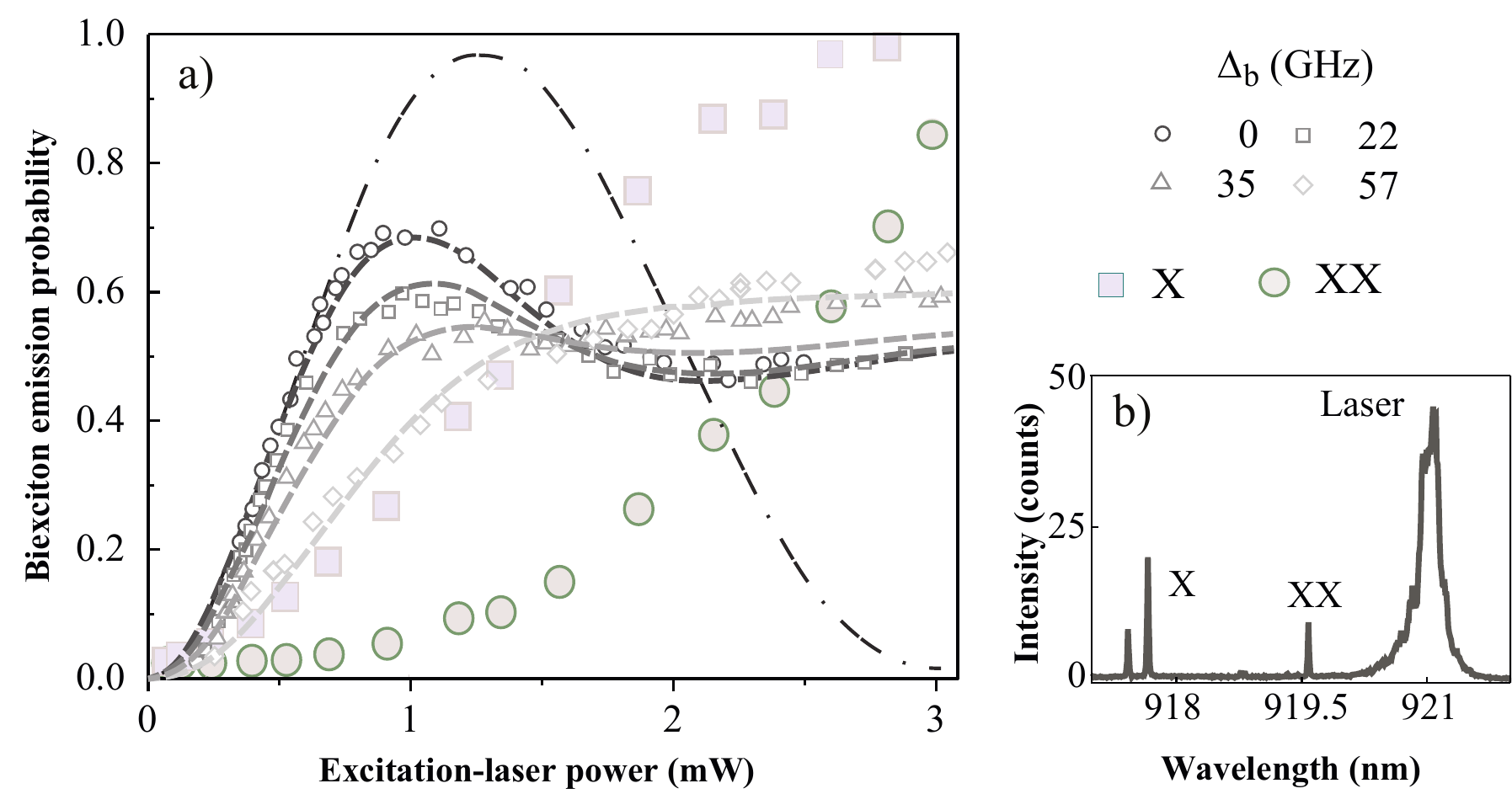}
\caption{(a) Rabi oscillations of the biexciton. The emission probability was measured for various excitation laser detunings $\Delta_{xx}$ from the two-photon resonance. The empty symbols represent the results of these measurements, circles, squares, triangles, and diamonds for $\big\{$0, 22, 35, 57$\big\}$ GHz detuning, respectively. Dashed lines are simulations employing the theoretical model described in \cite{Jayakumar2013}. This model includes additional dephasing processes that are not contained in the four Lindblad operator model described in this text. The model presented here results in the curve that is given as the dot-dashed line. The filled symbols stand for data obtained in power dependence measurement of biexciton and exciton photons under incoherent two-photon excitation that was performed using a laser detuned towards lower energy (red-detuned) for few nanometers from the two-photon virtual resonance. (b) Photo-luminescence signal obtained in this incoherent excitation regime.}
\label{fig:Rabi}       
\end{figure}

The Hamiltonian in matrix form is given as:

\begin{equation}
H=\hbar\left(\begin{array}{ccc}
0 & \frac{\Omega_{1}\left(t\right)}{2} & 0\\
\frac{\Omega_{1}\left(t\right)}{2} & -\Delta_{xx}+\Delta_{x} & \frac{\Omega_{2}\left(t\right)}{2}\\
0 & \frac{\Omega_{2}\left(t\right)}{2} & -2\Delta_{xx}
\end{array}\right).
\end{equation}
To calculate the state populations and corresponding emission probabilities we need to solve the master equation, here written in Lindblad form \cite{Carmichael1993,Gardiner2004}
\begin{equation}
\dot{\rho}=-\frac{i}{\hbar}[H,\rho]+\sum_{i=1}^{4}\mathcal{L}_{i}(\rho).
\end{equation}
Following \cite{Carmichael1993,Gardiner2004}, we use the following Lindblad operator

\begin{equation}
\mathcal{L}_{1}(\rho)=\frac{\gamma_{xx}}{2}
(2\sigma_{x,xx}\rho\sigma_{x,xx}^{\dagger}-\sigma_{x,xx}^{\dagger}\sigma_{x,xx}\rho-\rho\sigma_{x,xx}^{\dagger}\sigma_{x,xx})
\end{equation}
to describe the spontaneous decay from the biexciton to the intermediate exciton state and the operator
\begin{equation}
\mathcal{L}_{2}(\rho)=\frac{\gamma_{x}}{2}(2\sigma_{g,x}\rho\sigma_{g,x}^{\dagger}-\sigma_{g,x}^{\dagger}\sigma_{g,x}\rho-\rho\sigma_{g,x}^{\dagger}\sigma_{g,x})
\end{equation}

to describe the spontaneous decay from the exciton to the ground state. Unfortunately in quantum dots there are decoherence mechanisms that can put an end to Rabi oscillations before the spontaneous decay does so. The drift of the quantum dot energy levels is a well-known problem that impedes quantum dots from emitting Fourier transform limited photon wave-packets \cite{Santori2002}. Therefore it is essential to introduce Lindblad terms that describe the dephasing of the quantum dot levels due to its interaction with the environment. Again following \cite{Carmichael1993,Gardiner2004}, we can introduce the following Lindblad operators to model the dephasing of the biexciton level

\begin{eqnarray}
\mathcal{L}_{3}(\rho) &=& \frac{\gamma_{dxx}}{2}(2(\sigma_{xx,xx}-\sigma_{x,x})\rho(\sigma_{xx,xx}-\sigma_{x,x})^{\dagger}-\rho(\sigma_{xx,xx}-\sigma_{x,x})^{\dagger}(\sigma_{xx,xx}-\sigma_{x,x}) \nonumber \\
 &-& (\sigma_{xx,xx}-\sigma_{x,x})^{\dagger}(\sigma_{xx,xx}-{\sigma_{x,x}})\rho),
\end{eqnarray}

and respectively to describe the dephasing of the exciton level

\begin{eqnarray}
\mathcal{L}_{4}(\rho) &=& \frac{\gamma_{dx}}{2}(2(\sigma_{x,x}-\sigma_{g,g})\rho(\sigma_{x,x}-\sigma_{g,g})^{\dagger}-\rho(\sigma_{x,x}-\sigma_{g,g})^{\dagger}(\sigma_{x,x}-\sigma_{g,g})  \nonumber \\
 &-& (\sigma_{x,x}-\sigma_{g,g})^{\dagger}(\sigma_{x,x}-\sigma_{g,g})\rho).
\end{eqnarray}




Here, $\gamma_{xx}$ and $\gamma_{x}$ are the spontaneous decay rates and $\gamma_{dxx}$  and $ \gamma_{dx}$  are the dephasing rates of the biexciton and exciton, respectively. The excitation pulse is considered to have a Gaussian envelope function. 
Parameters like spontaneous decay and dephasing rates can be determined from experimental lifetime and coherence time measurements, respectively. Using these experimentally measured parameters we can numerically solve the master equation and thereby obtain the theoretical prediction for the populations of the different levels involved $(P_{i}=\langle \sigma_{ii} \rangle)$. The population multiplied with decay rate integrated over time gives the emission probability. The emission probability as a function of the square of the Rabi frequency in resonant excitation shows an oscillating behaviour commonly known as the Rabi oscillations.

The system studied in \cite{Jayakumar2013} showed measured lifetimes of $\tau_{xx}=1/\gamma_{xx} $ = 405~ps for the biexciton and $\tau_{x}=1/\gamma_{x} $ = 771~ps for the exciton. The coherence lengths of the emitted photons were measured to be $\tau_{dxx}=1/\gamma_{dxx}$  = 211~ps for the biexciton photon and  $\tau_{dx}=1/\gamma_{dx}$ = 119~ps for the exciton photon while the excitation pulse was measured to be 4~ps long. 
The theoretical prediction calculated for these parameters is given in Fig.~\ref{fig:Rabi}a as the dot-dashed curve, which indicates a very high emission probability at the adequate excitation strength. Such a result is not surprising because the excitation pulse length is much shorter than the dephasing mechanisms that were elaborated above. The source of lower than unity emission probability in this model can be attributed to the proximity of the two-photon virtual level to the exciton level ($ \Delta_{x}$=$2\pi~335$ GHz). Therefore, one can expect that in the absence of additional sources of dephasing such a photon source would create photon pairs on demand.

None the less, the experimental results shown in both \cite{Jayakumar2013} and \cite{Muller2014} show stronger dephasing of the Rabi oscillations. In \cite{Muller2014} this result was attributed to a chirp of the excitation pulse although authors did not exclude the existence of additional sources of dephasing. The findings given in \cite{Jayakumar2013} suggest the existence of an underlying incoherent process that dephases the excitation process. 
 In particular, in \cite{Jayakumar2013} it was shown that the photo-luminescence signal can be observed even when the quantum dot was addressed using a laser of an energy lower than the biexciton transition, Fig~\ref{fig:Rabi}b. The power dependence measured under these conditions showed that the exciton photo-luminescence signal increases quadratically with power, while the biexciton signal grows with fourth power. 
 The data obtained in these measurements are shown as full coloured symbols in Fig~\ref{fig:Rabi}a. While a two-photon process in the surrounding material (GaAs is highly nonlinear) that creates carriers in the vicinity of the quantum dot is possible, such a process would not cause the damping of the Rabi oscillation but rather a background in the photo-luminescence signal. On the other hand a process such as two-photon excitation from the ground state to the continuum would dephase the excitation process. 
 




\section{Measurements under resonant excitation}
\label{sec:coherent and time}

Here, we will briefly review several results that were obtained using resonantly excited quantum dots. In particular we will address the topics of  the coherent manipulation of the ground-excited state superposition, photon statistics of a resonantly excited quantum dot, and the generation of time-bin entangled photon pairs.

\subsection{Coherent control}
\label{sec:coherent}

\begin{figure}[t]
\centering
\includegraphics[scale=.46]{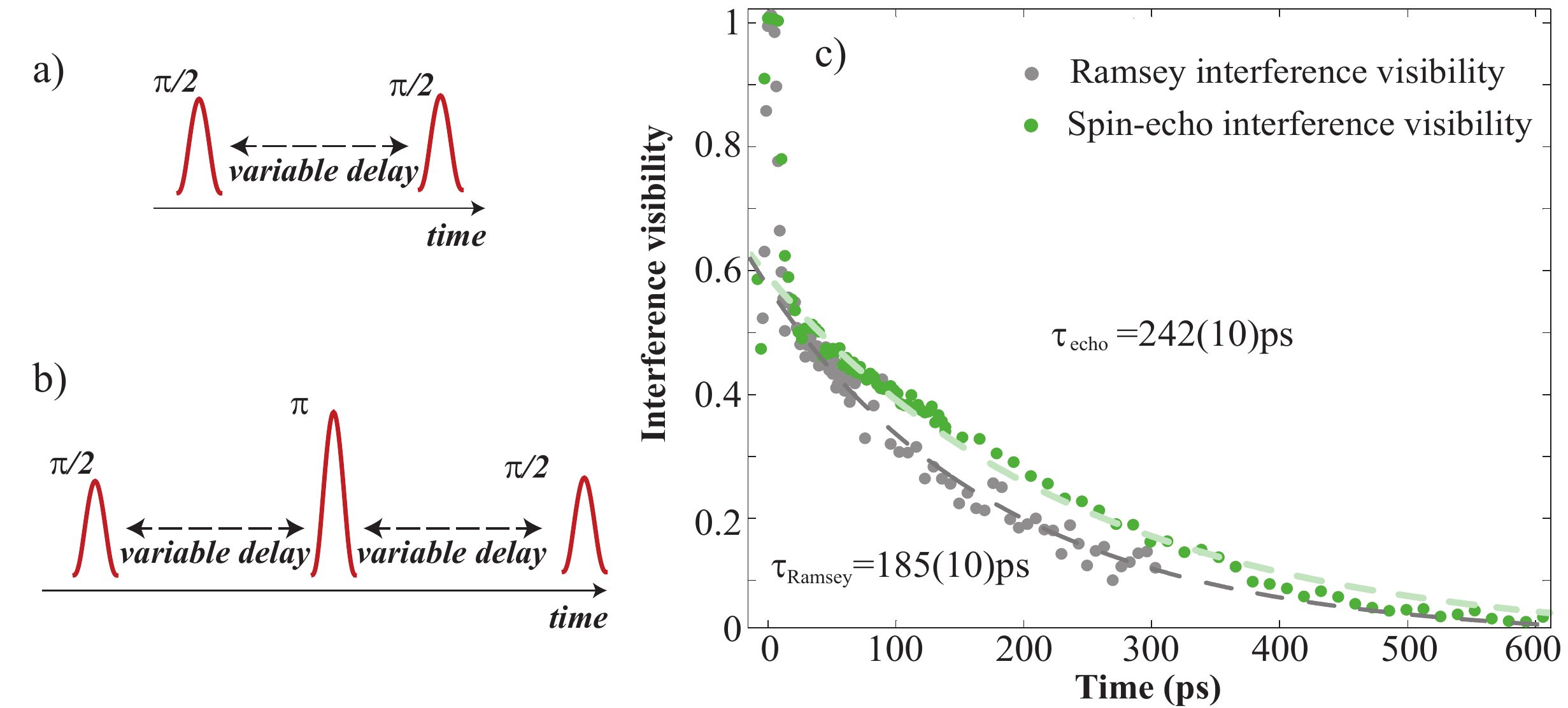}
\caption{(a) Pulse sequence consisting of two $\pi/2$ pulses applied with variable delay. (b) The spin-echo pulse sequence. (c) Ramsey interference visibility decay experiment monitored with the biexciton photons is shown in gray. The data shown in green were taken in a spin-echo measurement performed on the same emitter.}
\label{fig:Ramsey}       
\end{figure}

The \keyword{coherence} of the excitation process allows for the phase of the ground-biexciton state superposition to be coherently manipulated. The traditional way to characterize such a process is to perform a Ramsey interference measurement. To do so, one needs to excite the investigated system using a sequence of two consecutive $\pi/2$ pulses, Fig.~\ref{fig:Ramsey}a. The first of these pulses brings the state in an equal superposition of the ground and the biexciton state. 
Upon this pulse, one lets the system to evolve freely for a time defined by the variable delay between the pulses, Fig.~\ref{fig:Ramsey}a. During this free evolution the excitation pseudo-spin is expected to precess along the equator of the Bloch sphere. The second pulse will map the population either back to the ground state or flip it further to the biexciton state, depending on the evolution of the pseudo-spin and the relative phase between the two pulses. A very thorough review of the coherent manipulation of excitons and spins in quantum dot systems is given in \cite{Ramsay2010}.

When such an experiment is performed in two-photon excitation it results in Ramsey interference fringes in both the exciton and the biexciton emission \cite{Flissikowski2004}. It is important to note here that in the case of the biexiton emission these fringes are a direct result of the laser driving the transition. The interference observed in exciton channel closely follows the behaviour of the biexciton but comes as a consequence of the cascade decay of the system\footnote{Note that the Ramsey interference measurement characterizes the coherence of the ground-biexciton state superposition and that by varying the delay between the two Ramsey pulses one can measure the coherence decay of this pseudo spin.}. 
An example a decay of the Ramsey visibility fringes is shown in Fig.~\ref{fig:Ramsey}c. 

Decoherence caused by low frequency noise can be eliminated by applying a refocusing pulse. Such a measurement is commonly called \keyword{spin echo} and requires a sequence of three consecutive pulses of different intensities ($\pi/2, \pi, \pi/2$), illustrated in Fig.~\ref{fig:Ramsey}b. Due to their lifetime quantum dots are usually excited using laser pulses that are not longer than few picoseconds. Therefore the simplest way to obtain the sequence of Ramsey pulses is by feeding pulsed laser light into a variable-length Michelson interferometer. Concerning the spin echo measurements, it is quite straightforward to implement such a measurement in systems that have long lifetimes and coherence lengths. For example, for a trapped ions system where the coherences are of the order of a milisecond one can use light derived from a cw laser and create the pulse sequence using an acousto-optical modulator. Unfortunately, and as mentioned before, the pulse lengths needed to drive a spin-echo sequence on a pseudo-spin of a ground-biexciton state superposition of a quantum dot are on the order of few picoseconds. In \cite{Jayakumar2013} it was shown that the echo sequence with such pulses can be made by using a Michelson interferometer in  double-pass configuration. Such an implementation is capable of delivering the three consecutive pulses necessary for the spin-echo sequence with the middle pulse being a result of the interference between the light passing once through the interferometer with the light passing twice. In Fig.~\ref{fig:Ramsey}c are shown two sets of data, one taken in a Ramsey and the other in spin-echo experiment. We observe an increase in the visibility decay from $\tau_{Ramsey}$ = 185(10)~ps to $\tau_{echo}$ = 242(10)~ps. The measured values 
indicate the presence of high frequency noise, which could not be refocused by the spin echo technique. On the other hand, the technique itself is limited by the strong incoherent process that happens during the excitation and that is, as mentioned before, also responsible for the dephasing of the Rabi oscillations. 

\subsection{Photon statistics under resonant excitation}\

The statistics of the photons emitted by semiconductor quantum dots shows an intrinsically sub-Poissonian distribution \cite{Michler2000}. Individual emitters are commonly characterized by  a measurement of the \keyword{autocorrelation} \cite{Grangier1986} parameter (very often also called the $g^{(2)}(0)$ measurement). 
The choice of this particular method is historically rooted. The use of autocorrelation measurements on quantum dots can be traced back to the first experiments that were capable to address a single quantum dot and where the main experimental task was to isolate a single emitter from an ensemble of quantum dots. In such an experiment the observation of an autocorrelation parameter lower than 0.5 was a clear confirmation that the observed system was a single emitter.  While the autocorrelation is quite easy to implement experimentally from a certain perspective it is a limited resource. Namely, it is an efficiency insensitive measurement that alone cannot deliver the absolute values for the photon generation probabilities $p_1$ or $p_{2+}$ (probability for a single photon and multiple photons, respectively). On the other hand, and in the limit of the low source efficiency, the autocorrelation parameter can be approximated as 2$p_{2+}$/($p_1$+2$p_{2+}$)$^{2}$, \cite{gauss}.

\begin{figure}[t]
\centering
\includegraphics[scale=0.63]{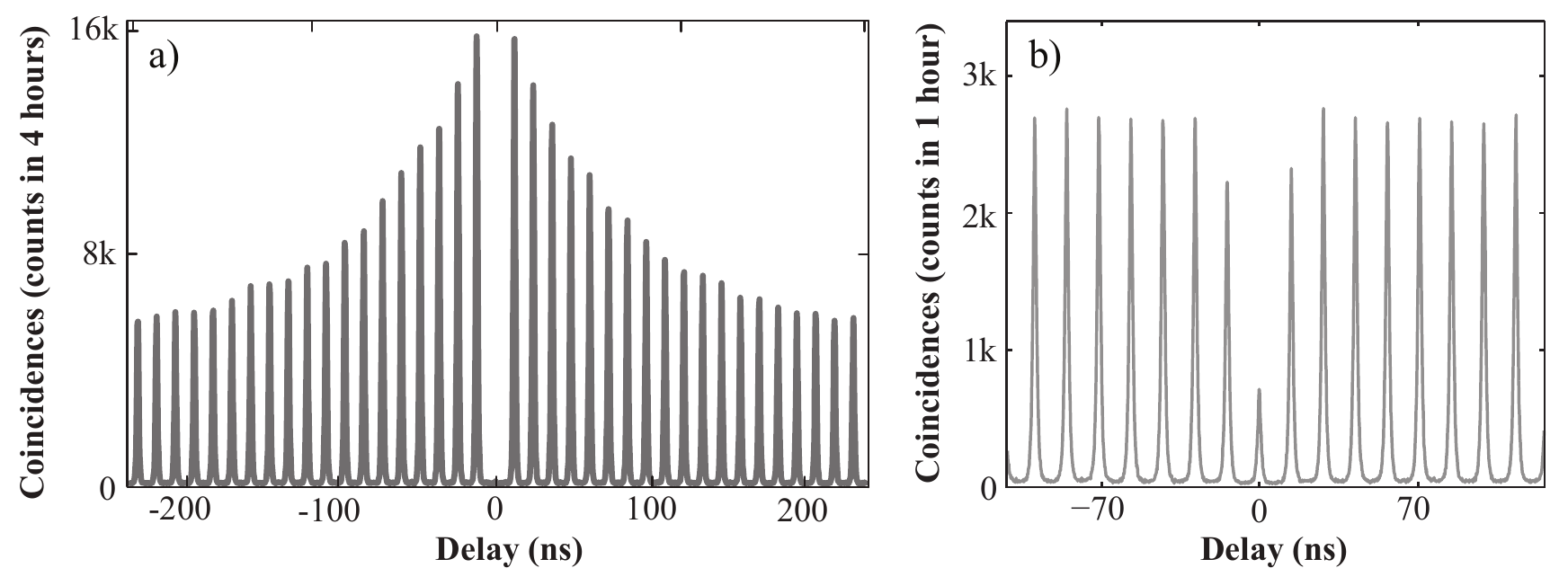}
\caption{(a) The exciton signal shows excellent suppression of multi-photon events which can be quantitatively expressed by intensity autocorrelation parameter of 
0.0315(2). The plotted data is presented without background subtraction. The decaying peak height observable on both sides of the graph results from the blinking of the quantum dot \cite{Santori2004}. (b) The same quantum dot will show far larger probability of multiple excitations when excited above-band.}
\label{fig:hbt}      
\end{figure}

Today the problem of addressing a single isolated quantum dot can be considered no longer challenging and the attention is redirected to the increase of the collection efficiency and reduction of the \keyword{multi-photon component}. The latter, in the case of quantum dots, can be reduced to the problem of multiple excitations. As mentioned before, the traditional way to excite quantum dots is above-band excitation. Apart from a lack of coherence in driving the quantum dot system this excitation method also gives probabilistic statistics for the photon generation and can induce effects that increase the multi-photon component in the statistics of the emitted light like carrier re-capture \cite{Peter2007}. The latter is well illustrated in Fig.~\ref{fig:hbt}. The autocorrelation measurement given in Fig.~\ref{fig:hbt}a was made on resonantly excited quantum dot, and the autocorrelation parameter extracted from the data reads 0.0315(2). On the other hand, the same quantum dot excited in above-band excitation will show much higher multi-photon component, shown in Fig.~\ref{fig:hbt}b. Here the the autocorrelation parameter reads 0.282(1).

\begin{figure}[t]
\centering
\includegraphics[scale=0.45]{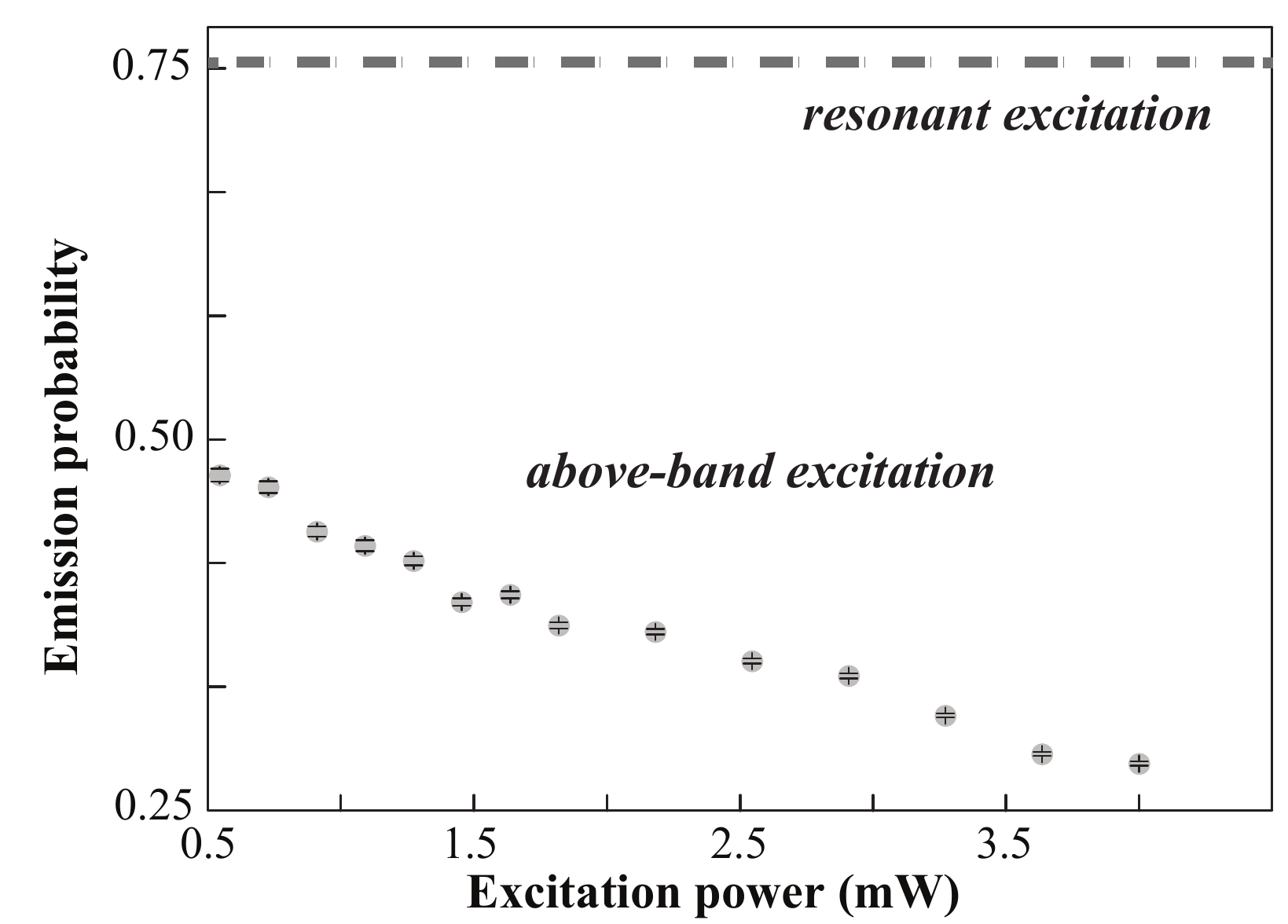}
\caption{Here, the dashed line marks the maximum emission probability of an biexciton-exciton photon pair obtained under two-photon resonant excitation. The gray circles show the same probability under above-band excitation. For the latter measurement the excitation power up was gradually increased up to the level where the biexciton saturates (4~mW).}
\label{fig:eff}      
\end{figure}

An autocorrelation measurement is not sufficient to recognize the efficiency of the emitter nor the relation between the efficiency and the multi-photon component of the emission. Nevertheless, there are a number of measurements that one can perform and obtain these numbers including \cite{gauss, onion, Filip2011, Jezek2011}. Concerning the efficiency alone, the resonant excitation will always give better numbers, as illustrated in Fig.~\ref{fig:eff}. Here, the saturation of the biexciton level demands large concentration of carriers in the quantum dot surrounding. This implies a large probability that after the biexction photon has been emitted the system never reached the ground state but rather immediately captured another electron-hole pair.

\subsection{Time-bin entanglement}
\label{sec:time-bin}

\begin{figure}[b]
\centering
\includegraphics[scale=1]{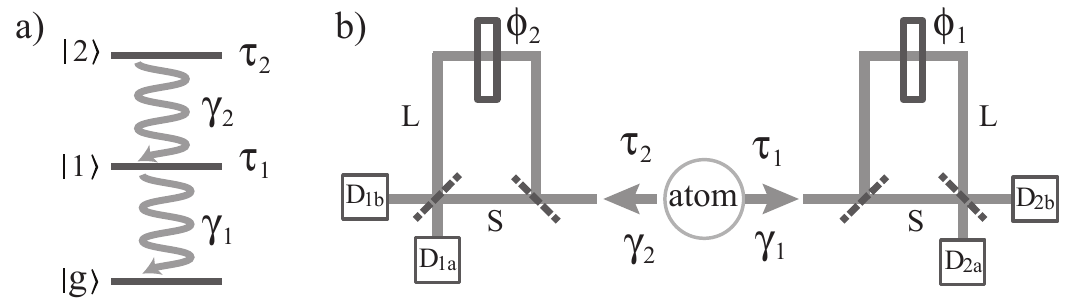}
\caption{ (a) Three-level atomic system with a long lifetime $\tau_2$ for the initial state and a much shorter lifetime $\tau_1$ for the intermediate state. The decay rates are denoted $\gamma_2$ and $\gamma_1$, respectively. (b) The state-analysis interferometers with the short (S) and long (L) path. In the absence of the interferometers (or interferometer beamsplitters) the coincidence measurement between the $D_1$ and $D_2$ detectors would show a very narrow peak of width $\tau_1$. Nevertheless, the uncertainty for the photons to be emitted was initially much longer ($\tau_2$) and therefore the associated wave packet must have had large time and position uncertainty. The detection of one of the photons has as an effect a nonlocal change in the wavefunction describing the other photon. By varying the relative phase between the interferometers ($\phi_2-\phi_1$) one can observe the visibility contrast and therefore measure the entanglement.}
\label{fig:franson}      
\end{figure}

The idea to generate \keyword{time-bin entanglement} of photons emitted by a single quantum emitter can be traced back to a seminal paper by J.D. Franson \cite{Franson1989}. He suggested that the interference between the probability amplitudes for a photon pair to be emitted by an excited atom at diverse times is a nonlocal effect that violates the Bell inequality. The system described by Franson consists of an atom in an excited state that decays to the ground state via emission of a photon cascade (pair of photons), Fig~\ref{fig:franson}a. The necessary condition given in this proposal is that the atom has a very long-living initial state. Additionally, the intermediate excited state needs to be very short-lived so that the second photon of the cascade is emitted immediately after the first photon. The interference is observed in coincidence events and its detections employs two unbalanced interferometers, one for the each photon of the cascade, Figure \ref{fig:franson}b. The imbalance of the interferometers is supposed to be longer than the coherence length of the emitted photons otherwise the oscillations in the detected signal come from the beat of the field function.

The first implementations of Franson's scheme were elaborated using spontaneous parametric down-conversion \cite{Ou1990,Kwiat1990}. These experiments used a narrowband continuous wave laser to produce a pair of photons highly correlated in frequency and sent them down a pair of unbalanced interferometers. Here, the role of the long-lived excited state is played by a highly coherent monochromatic laser. These experiments were very challenging and suffered from several technological shortcomings such as poor detector resolution. Nevertheless, once these difficulties were overcome, such a type of an experiment showed violation of the Bell inequality \cite{Kwiat1993}.

The experiments using narrowband continuous wave lasers were producing photons that were time-energy entangled. In 1999, Brendel et al. \cite{Brendel1999} introduced a scheme that employs femtosecond-laser pumped parametric down-conversion. Such a scheme is commonly denoted as time-bin entanglement.  Here, instead of continuous wave long-coherence laser, one uses a pulsed laser of short coherence. The light derived from such a laser is sent into an unbalanced interferometer. Each laser pulse gives two pulses at the exit of the interferometer, so-called early and late pulse. These are directed onto the system we want to excite and if the excitation probability is kept low the system will on average be excited by only the early or only the late pulse. The state analysis is performed in a similar manner as proposed by Franson\footnote{At this point, it becomes obvious that the coherence properties of the pump laser have a very important role in the generation of the entanglement. An intermediate regime of a continuous wave short coherence pump laser was investigated in \cite{Liang2011}. They showed that the short coherence of the pump laser limits the visibility contrast to 50$\%$ }.

Therefore, time-bin entanglement 
encodes quantum states in superposition of the system's excitation within two distinct time-bins: early and late. The importance of this type of encoding lies in optical-fibre based quantum communication \cite{Honjo2007,Dynes2009}, due to the degradation that polarization entanglement can suffer in an optical fibre outside laboratory conditions \cite{Brodsky2011}. The issue behind the degradation of the polarization entanglement in optical fibres is polarization mode dispersion. It is a problem well known in telecommunication technologies that limits the rate of the information transfer. The physical origin of this effect is that even single mode fibres have two well-defined and differentiated polarization modes that in absence of a controlled environment will couple randomly. In addition to being insensitive to polarization mode dispersion and therefore preferred for a fiber optics long distance communication protocols this type of entanglement can also be employed in quantum computing. Recently a method was demonstrated to perform linear optical quantum computing using photons entangled in time-bin \cite{Humphreys2013}.

\begin{figure}[b]
\centering
\includegraphics[scale=1.15]{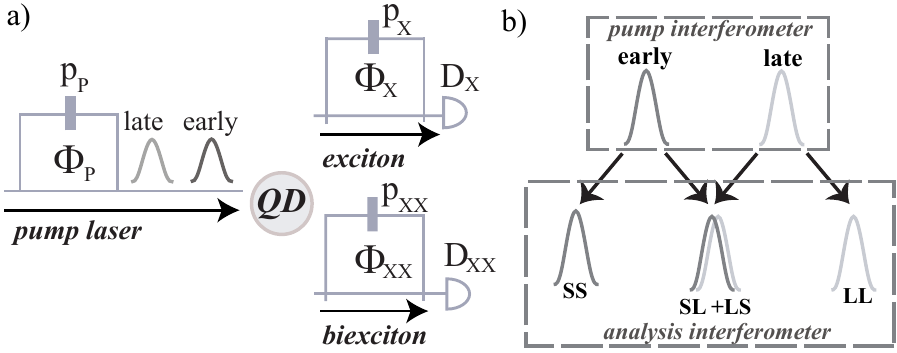}
\caption{ (a) Schematic of time-bin entanglement. The quantum dot (QD) is excited by two consecutive pulses derived from an unbalanced Michelson interferometer shown on the left. The interferometric phase between these pulses is $\phi_P$. The state analysis is performed using another two unbalanced interferometers, one for exciton and other for biexciton photons. These two interferometers have their respective phases, $\phi_X$ and $\phi_{XX}$. The photons are detected upon leaving the analysis interferometers using detectors $D_X$ and $D_{XX}$. The phases of the individual interferometers are controlled using phase plates, $p_P$, $p_X$ and $p_{XX}$. (b) A photons pair created by an early pulse but later in analysis travelled the long paths of the analysis interferometers is in its arrival time indistinguishable to a photon pair created in a late pulse that in analysis travelled the short paths.}
\label{fig:time-bin}      
\end{figure}

In its simplest scheme, time-bin entanglement is generated in a very similar manner for both parametric down-conversion \cite{Brendel1999} and atom-like systems \cite{Jayakumar2014} and it demands post-selection in order to be measured. Such a scheme \footnote{A quantum dot system can give time-energy entangled photon pairs, nevertheless it needs independent control over the lifetimes of the involved energy levels \cite{Huber}} is depicted in Fig.~\ref{fig:time-bin}(a). The system is addressed by two excitation pulses, denoted the early and the late pulse. These are derived from an unbalanced interferometer, so-called pump interferometer. The interferometric phase, $\phi_{P}$, between the pulses determines the phase of the entangled state. The analysis of the generated state is performed using two identically-constructed unbalanced interferometers, one for exciton and one for biexciton photons. The entangled state reads

\begin{equation}
|\Phi\rangle=\frac{1}{\sqrt2}(|{\rm early} \rangle_{XX}|{\rm early} \rangle_{X}+e^{i\phi_{P}}|{\rm late}\rangle_{XX}|{\rm late}\rangle_{X}),
\label{state}
\end{equation}

where $\phi_{P}$ is the phase of the pump interferometer and $|{\rm early}\rangle$ ($|{\rm late}\rangle$) denote photons generated in an early (late) time-bin.
The method to write the phase, $\phi_{P}$, onto the system differs between parametric down-conversion and atom-like systems. In particular, quantum dots demand resonant excitation in order to bring the system from the ground to the excited state coherently, while in the process of parametric down-conversion the phase matching process itself ensures that the pump laser and the down-converted fields maintain constant phase relation.

The factors limiting the levels of time-bin entanglement obtainable from an atom-like system are of two types: excitation-method specific and system-coherence related. The first ones can be rendered to so-called double excitations. In particular, if the entanglement measurement was performed with excitation probability $p_1$, it will happen in $p_1$\textsuperscript{2} cases that the system is excited by both the early and the late pulse. These events are observable in time basis\footnote{${\rm early}/{\rm late}(E/L) $} being less than unity, and are as well present as incoherent background in both energy bases\footnote{$E+L/E-L$ and $E+iL/E-iL$}. The effect of double excitations can be eliminated through use of deterministic schemes for generation of time-bin entanglement \cite{Simon2005,Pathak2011,Nisbet2013}.

The time basis measurements are not affected by the decoherence-induced reduction of the visibility contrast; in contrary the energy bases measurements are.  An intuitive picture of how the decoherence affects the time-bin entanglement is the following: the pump interferometer phase, $\phi_{P}$, is transferred onto the quantum dot by means of resonant excitation. Any incoherence in the process of resonant excitation as well as in relation between the ground and the biexciton state will have as a consequence an averaging of the transferred phase, and thus of the phase of the entangled state. This will reduce the visibility contrast as well as decrease the values of entanglement measures and indicators like concurrence and fidelity. 

It is important to notice that the entanglement analysis involves the two interferometers depicted in Fig.~\ref{fig:time-bin}(a) (one per qubit) and is a method that includes post selection. The post-selection procedure is schematically plotted in Fig.~\ref{fig:time-bin}(b). Namely, the emission time of photons contains the information on which pulse has created the photon pair and the analysis interferometers can partially erase this information. In particular, a photons pair that was created by an early pulse and in analysis travelled the long paths of the interferometers is in its arrival time indistinguishable to a photon pair that was created in a late pulse and in analysis travelled the short paths. If the detectors and the analysis electronics are fast enough to isolate these indistinguishable events the entanglement can be measured.

\begin{figure}[t]
\centering
\includegraphics[scale=.30]{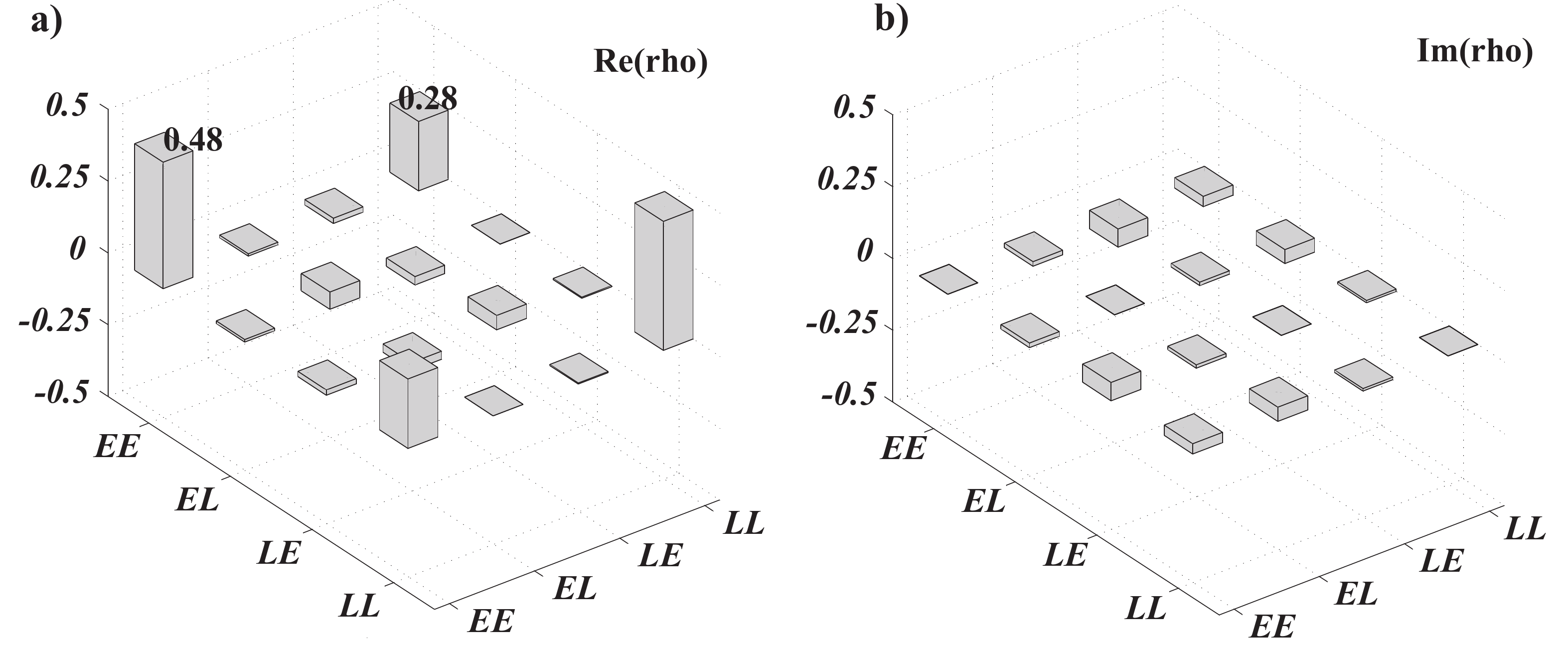}
\caption{An example of a (a) real and (b) imaginary part of the reconstructed density matrix. Measurements used to obtain this density matrix were performed using 4~ps long excitation pulses while the excitation probability was kept at 6$\%$}
\label{fig:matrix}      
\end{figure}

\begin{figure}[b]
\centering
\includegraphics[scale=.6]{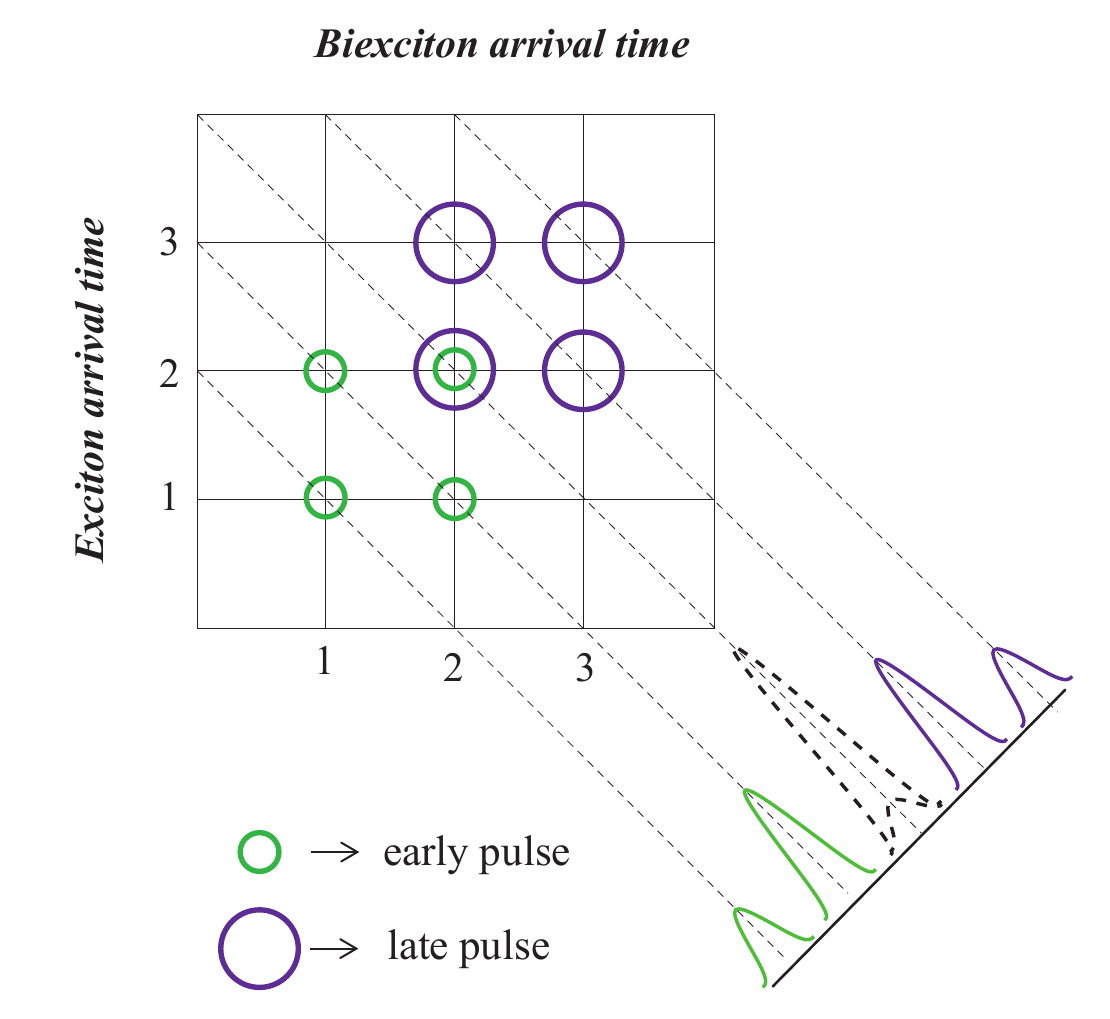}
\caption{Time-bin entanglement is observed in the coincidence events. The plot depicts the arrival times of the individual photons and how these are forming the coincidence counts. The green circles stand for arrival times of the photons produced by an early pulse while the violet circles stand for arrival times of the photons produced by a late pulse. Interference of the probability amplitudes is observed at the point of overlap. The diagonal projection shows a coincidence pattern. The existence/absence of the central peak, plotted in black, shows the coincidence correlation/anti-correlation.}
\label{fig:5peaks}      
\end{figure}

A very complete method to characterize the entanglement is state tomography \cite{James2001,Takesue2009}. Fig.~\ref{fig:matrix} shows an example of a reconstructed density matrix. The fidelity of this particular matrix with the maximally entangled state was found to be F=0.78(3) while the tangle and concurrence are T=0.31(9) and C=0.56(7). 

In addition, the measurement shown in Fig.~\ref{fig:matrix} also gave the visibilities in three orthogonal bases of 92(2)$\%$, 52(3)$\%$, and 57(3)$\%$ for $E/L$, $E+L/E-L$, and $E+iL/E-iL$, respectively. It is worth mentioning that among the measurements that can indicate entanglement the visibility is the simplest and probably the oldest. This particular measurement was forming the central part of the original Franson's proposal \cite{Franson1989} and also was the method used in the first measurements of the time-energy entanglement \cite{Ou1990,Kwiat1990}. It is observed in coincidental events. Figure~\ref{fig:5peaks} shows the diagram of arrival times of the two photons in the time-bin measurement and the corresponding coincidence peaks. The central peak comes from the interference of the probability amplitudes.  If the system is driven with too high excitation probability it leads to increase in number of events where both the early and the late pulse have generated an excitation. It is clear from this plot that such events will form a background under the central coincidence peak that reduces the maximum achievable visibility.

\section{Future directions}
\label{sec:conclusion}

Quantum dots are systems that show great potential. They are compact and integrable in solid state devices. When driven resonantly their atom-like nature allows for high photon generation probability complemented with low probability for multiple excitations. With respect to on-demand generation of photon pairs, the two-photon resonant excitation is a very promising method.

On the other hand, our knowledge on the origin and the nature of decoherence processes in quantum dots is still scarce. The ability to excite the quantum dot resonantly, to coherently manipulate the ground-excited state superposition, and/or to generate time-bin entanglement open up a possibility to use these measurement to further study and characterize the origins of decoherence.
 
One topic that was not addressed within this chapter is the extraction efficiency. Namely, the extraction efficiency in samples with planar micro-cavities is higher than in the dots without any additional structure but is still limited to about $5\%$ in best cases. Using etched micro-pillar cavities dramatically increases the collection efficiency. The two-photon resonant excitation can readily be applied to these devices too. We expect the benefits to be multi-fold, ranging from the increased collection efficiency to Purcell enhancement of the emission and therefore a emission of a shorter and less decohered wave-packet. Such an approach would allow for having a photon source capable of fulfilling the high requirements set by quantum information science protocols and schemes.

\section{Acknowledgements} %
\begin{acknowledgement}
{I would like to express my gratitude to all my colleagues and collaborators, that over last couple of years, have taken part in the experiments conducted 
at the University of Innsbruck. Writing of this review was supported by the Austrian Science Fund (V-375). This review in part presents the results of research that was financially supported by Austrian Science Fund (M-1243 and V-375) and University  of Innsbruck.}
\end{acknowledgement}

\bibliographystyle{SpringerPhysMWM} 

\bibliography{Example}

\printindex

\end{document}